\documentclass{article}
\usepackage{spconf,amsmath,graphicx}
\usepackage{booktabs}
\usepackage{adjustbox}

\title{Self-supervised adaptive AV Fusion Module for pre-trained ASR models}
\name{Christopher Simic, Tobias Bocklet}
\address{Technische Hochschule Nürnberg Georg Simon Ohm, Germany}
\begin{document}
\ninept
\maketitle
\begin{abstract}

Automatic speech recognition (ASR) has reached a level of accuracy in recent years, that even outperforms humans in transcribing speech to text. Nevertheless, all current ASR approaches show a certain weakness against ambient noise. To reduce this weakness, audio-visual speech recognition (AVSR) approaches additionally consider visual information from lip movements for transcription. This additional modality increases the computational cost for training models from scratch. 
We propose an approach, that builds on a pre-trained ASR model and extends it with an adaptive upstream module, that fuses audio and visual information. Since we do not need to train the transformer structure from scratch, our approach requires a fraction of the computational resources compared to traditional AVSR models. 
Compared to current SOTA systems like AV-HuBERT, our approach achieves an average improvement of 8.3\,\% in word error rate across different model sizes, noise categories and broad SNR range. The approach allows up to 21\,\% smaller models and requires only a fraction of the computational resources for training and inference compared to common AVSR approaches.

\end{abstract}

\section{Introduction}
\label{sec:intro}
Automatic Speech Recognition (ASR) describes methods used to convert speech into text. Current ASR systems are trained on huge amounts of speech data and have reached a level that surpasses human accuracy in speech to text transcription tasks~\cite{INTRO_superHumPerf_2020}. In recent years, the use of transformer-based approaches has been established. These models consist of an encoder-decoder structure for embedding the input audio information within the encoder, while the decoder processes these information to generate the output text. Currently, the most common used approaches of this type are Whisper~\cite{ASR_whisper_2022}, HuBERT~\cite{ASR_HuBERT_2021} and wav2vec2~\cite{ASR_wav2vec_2020}. These differ in terms of input representation of audio information and loss calculation. While HuBERT and wav2vec2 use raw audio signals directly, Whisper expects log-mel spectrum inputs. The optimization of wav2vec2 is performed based on a Connectionist Temporal Classification (CTC) Loss calculation, while Cross Entropy (CE) Loss is used for HuBERT and Whisper. In addition to transformer-based architectures, conformer architectures~\cite{ ASR_conformer_2020}, that combine the advantages of transformers and CNNs, have also shown promising results.
All ASR approaches have in common that high ambient noise leads to a decrease in recognition accuracy. 
For humans it is easier to understand speech in noisy environments, if we can see the speakers face and interpret the movements of lips, teeth and tongue.
A similar idea is pursued by approaches to audio-visual speech recognition (AVSR), in which audio and visual information are processed together for transcription. 
Due to the correlation between lip movements and speech signal, the analysis of mouth shapes is the basis of many speech-based tasks such as single modal visual speech recognition (VSR) ~\cite{VSR_spattemp_fusbased_2019, VSR_relaxedAtt_2022}, speech enhancement~\cite{SEH_kalm_2023, SEH_AVSE_obstr_2019}, speech separation~\cite{SSEP_mm_ssup_emb_2023, SSEP_visvoice_2021}, speech synthesis~\cite{SSYN_vocSynth_2020, SSYN_l2aufsp_2018} and speaker verification~\cite{SPVER_CMAV_tispver_2023}.
In this work, we focus on the consideration of lip movements for audio-visual speech recognition (AVSR). 
A common feature of all AVSR approaches is the use of CNNs to pre-process image-based information on lip movements (visual features). First promising results for processing visual features to improve speech recognition tasks have shown recurrent networks like LSTM~\cite{AVSR_LipR_Profile_2017}, GRU~\cite{AVSR_GRU_2018} or RNNs~\cite{AVSR_rnnt_AVSR_2019, AVSR_discr_MMSR_2020}. Most recently, the introduction of transformer-based~\cite{AVSR_deep_AVSR_2018, AVSR_jointly_rawD_2022} and conformer-based~\cite{AVSR_ete_AVSR_conf_2021, AVSR_autoAVSR_2023} architectures, analogous to the development in ASR led to a significant improvement in recognition rates. State of the art approach for AVSR is currently AV-HuBERT~\cite{AVSR_avhubert_2022, AVSR_robust_ssAVSR_avhubert2_2022}, which performs audio feature extraction through a feed forward network (FFN) and visual feature extraction through a CNN ResNet. The feature vectors of both modalities are then concatenated and processed through an encoder-decoder transformer system.
During pre-training, the encoder is optimized on the result of an acoustic clustering. The finetuning of encoder and decoder takes place afterwards, on a reduced dataset. Since current transformer-based approaches are trained from scratch, high computational ressources are required for training. 

We present an approach that takes advantage of a pre-trained ASR model and extend it with an upstream module that considers visual information of lip movements to increase robustness against ambient noise. For this purpose, after audio and visual feature extraction, a fusion of both feature vectors is performed using cross-modal attention, which results in a kind of enhancement of the ASR audio inputs. Because we adapt our fusion module to a pre-trained ASR model, which remains frozen in the first step, our approach requires significantly less computational resources than comparable AVSR approaches trained from scratch. We train our module in a self-supervised manner, which also provides the possibility to use unlabeled data for training. For this purpose, target data are generated at different model levels for clean inputs using the same ASR model. Finally, we finetune our fusion module and the ASR model to estimate the full potential of our approach.

\section{Data}
\label{sec:data}

For training and testing AVSR models, speech data with aligned video recordings of the speaker are required. For this purpose we use the LRS3-Ted~\cite{DS_LRS3} dataset, which is the standard dataset for AVSR approaches. LRS3-Ted consists of 433 hours of video material of around 152.000 short sequences from YouTube Ted talks. The dataset includes examples from more than 9000 speakers and a vocabulary of more than 51000 words. This makes the dataset significantly larger and more demanding than previous datasets. 

We focus on evaluating speech recognition under noisy conditions. Therefore, we need data with recordings of different noise types in order to specifically contaminate speech data and simulate noisy environments. For this purpose we use the Musan~\cite{DS_musan2015} dataset, that consists of audio sequences from the categories speech, music and natural sounds. Speech examples with a length of over 60 hours, music with 42 hours and natural noise of approximately 6 hours are available. The speech examples are divided into government speeches and reading recordings, the music part consists of music clips from various categories with and without vocals and the natural part consists of technical noise and ambient noise examples.

\section{Method}
\label{sec:method}

\subsection{Adaptive AV Fusion Module}
\label{ssec:adaptFusion}

\begin{figure}[htb]
\begin{minipage}[b]{1.0\linewidth}
  \centering
  \centerline{\includegraphics[width=8.5cm]{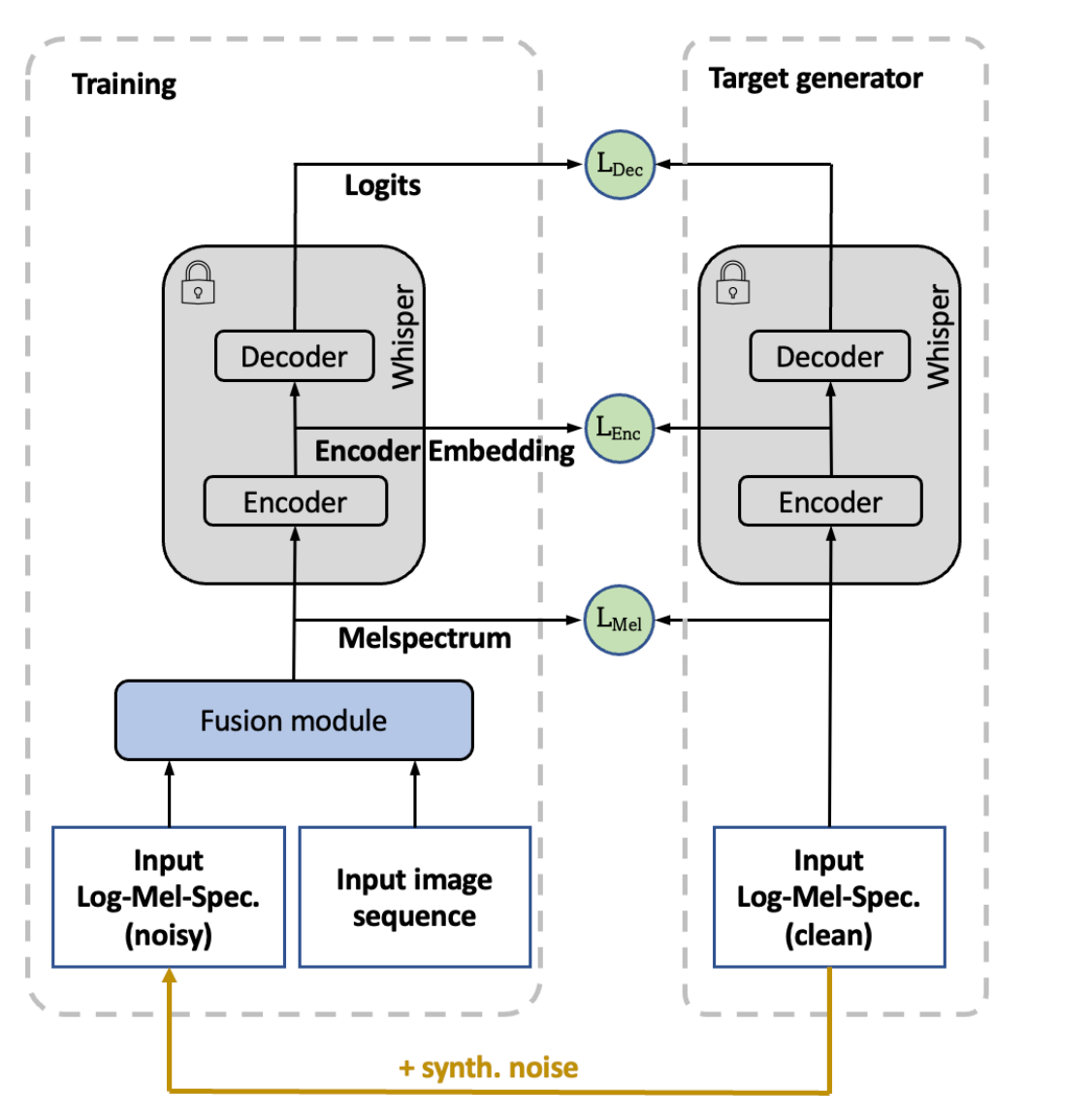}}
\end{minipage}
\caption{Overall model setup. The training setup (l) consists of the combination Whisper ASR model with our fusion module and training target generation (r)}
\label{fig:CompleteModel}
\end{figure}

Our fusion module can be trained in combination with any ASR model. In this work, we choose Whisper~\cite{ASR_whisper_2022} ASR, being one if the SOTA approaches trained on huge datasets, with a certain robustness against ambient noise~\cite{MT_whisperAT_2023}.
Whisper expects log-mel spectra of audio sequences as input and produces logit outputs, which are interpreted as tokens and thus text. 
Fig. \ref{fig:CompleteModel} shows our overall model setup. The trained fusion module receives the log-mel spectrum of a noisy audio signal and a corresponding video sequence of lip movements and produces vectors of the same shape as the input log-mel spectrum, which are fed into the ASR model.

\begin{figure}[htb]
\begin{minipage}[b]{1.0\linewidth}
  \centering
  \centerline{\includegraphics[width=8.5cm]{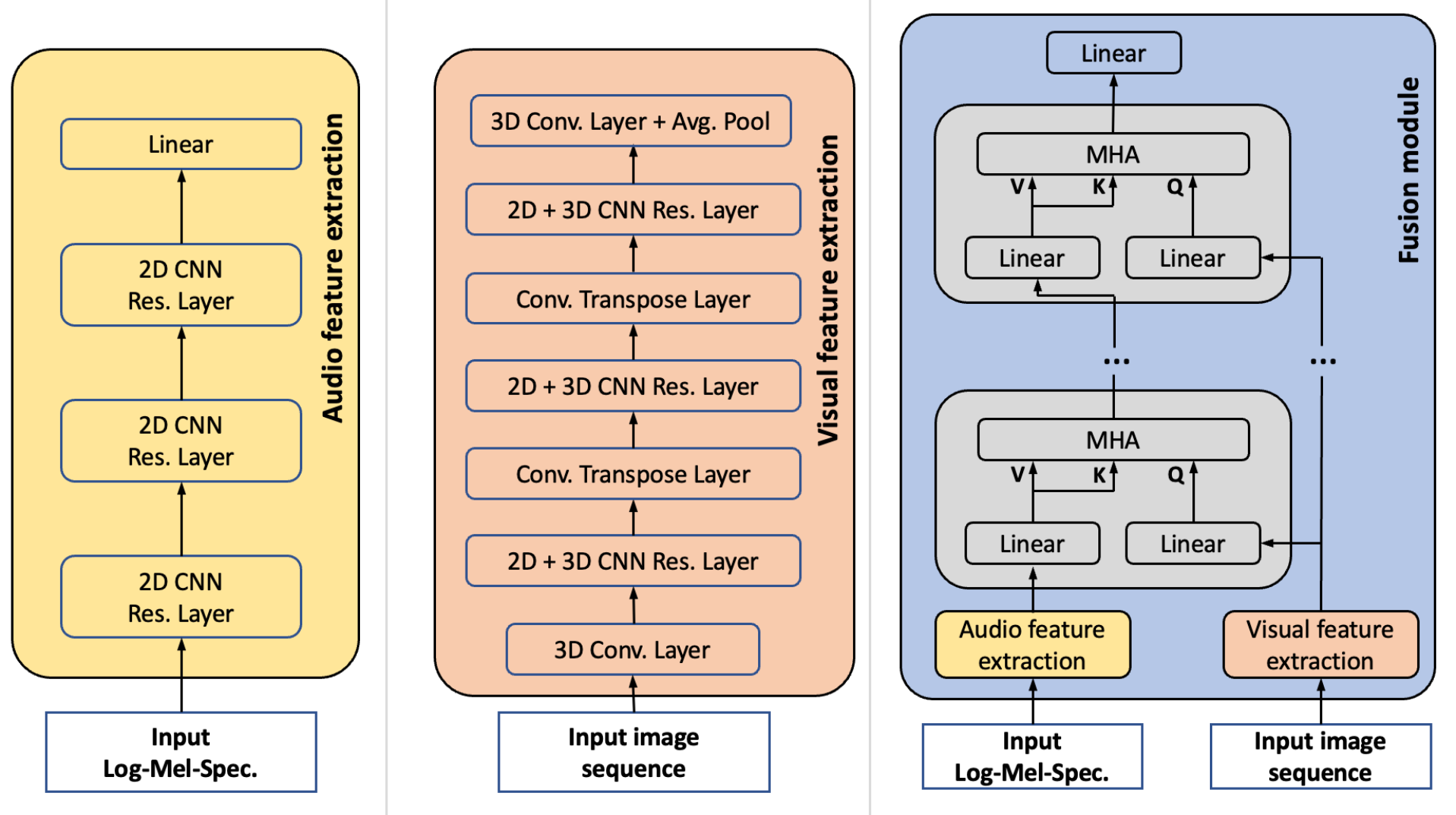}}
\end{minipage}
\caption{Our Fusion Module (r), consisting of two CNN based feature extraction models (details l, m) and a sequence of multihead cross-attention layers.}
\label{fig:Fusionmodule}
\end{figure}

The structure of our fusion module is shown in Fig. \ref{fig:Fusionmodule}.
Audio and visual inputs are processed by different CNNs. 
For audio processing we select a relatively simple 3 layer 2D CNN residual network. Each residual block consists of 2 convolutional layers with filtersize 3x3. To process the visual inputs, we use an alternating sequence of 2D and 3D residual blocks. As the frequency of the input log-mel spectrum vectors and video frames differ, our model requires two intermediate convolutional transpose layers to increase the visual feature output frequency by a factor of 4. A final convolutional layer with average pooling ensures that audio features and visual features are of the same shape before being processed by the fusion block.
The fusion of both modalities is performed in a sequence of multi-head cross-modal attention blocks. The first attention block takes the audio features as Value and Key, while the visual features are taken as Query. In later blocks, we continue to use the visual features as Query and the outputs of the previous block as Value and Key. In this way, our model executes an enhancement for the noisy input log-mel spectrum. The module consists of 12 attention layers with 12 attention heads each.

Fig.~\ref{fig:CompleteModel} illustrates the self-supervised character of our approach. For each training example, target data are generated from the clean audio signal at different positions of our model, which serve as optimization targets, while the same clean audio signal is synthetically contaminated by a noise signal and serves as input for the fusion module.

$L_{mel}$ calculates the loss on spectrum level, L\textsubscript{enc} on encoder embedding level and $L_{dec}$ on logit level after the ASR decoder. The ASR model remains frozen during first training steps. $L_{mel}$ and $L_{enc}$ calculate a $L_{1}$-loss, while a cross entropy (CE) loss is calculated for $L_{dec}$.

\begin{equation}
    L_1 = {|}y_{target}-y_{pred}{|}
    \label{eq:L1}
\end{equation} \\
The $L_1$-norm, which is the basis of the $L_1$-loss (\ref{eq:L1}), computes the sum of the absolute distances between each element of target data y\textsubscript{target} and prediction y\textsubscript{pred}.

\begin{equation}
    CE = -log \frac{exp(y_{\hat{c}})}{ \sum_{c=1}^C exp(y_{c})}
    \label{eq:Lce}
\end{equation} \\
For the CE-loss calculation (\ref{eq:Lce}) we use a slightly modified version. First, the logits from the target generator are one-hot encoded, to determine the target class $\hat{c}$. The value of the prediciton at the target class position $y_{\hat{c}}$ and the sum over all classes $C$ of a prediction are processed by an exponential function, before the loss function \textit{CE} computes the ratio of these values and aims to maximise its log value.

During training, all three losses $L_{mel}$, $L_{enc}$ and $L_{dec}$ are calculated separately and cumulated on a weighted basis. We choose a weight factor of 1.0 for $L_{mel}$ and 2.0 for $L_{enc}$ and $L_{dec}$, to set more focus on the results after ASR encoder and decoder during training. We tested different weight distributions to evaluate the most suitable combination.

\subsection{Model sizes and Training settings}
\label{ssec:ModelSizes}
Whisper offers different model variants of different sizes with 39\,M to 1550\,M parameter, while AV-HuBERT is available in two versions - base (103\,M) and large (325\,M). Our fusion module consists of 13\,M trainable parameter. In order, to have a fair comparison of our approach, combined with Whisper ASR models to AV-HuBERT base and large we select two Whisper variants, base.en (74\,M parameter) and small.en (244\,M).
We train and test our model on the LRS3-Ted dataset, of which we use the official pretrain split (400h) to pre-train our model, while optimizing only on the losses $L_{mel}$ and $L_{enc}$. For the main-training and finetuning, considering all three losses, we chose the trainval split (30h), with an additional 90:10 split for validation purposes. Accordingly, we select the AV-HuBERT models, that are fine-tuned on the 30h split for comparison.
Data preparation is performed analogous to AV-HuBERT. 96x96 pixel lip-centered frames are cropped from the video streams, with a randomly chosen area of 88x88 pixels during training. The input frames are converted to grayscale.

For synthetic noise generation, we chose the Musan dataset which is also prepared analog to AV-HuBERT, with different noise splits for training, validation and testing. Beside the Musan noise categories babble, music and natural (noise), we randomly select single sidespeaker sequences from the LRS3 dataset as additional noise category. During training, we define uniform distributed sound noise ratio (SNR) values from range [-20\,dB, +50\,dB]. In contrast to AV-HuBERT we ignore pauses in the audio sequences for SNR and scaling factor calculation. This leads to an relative SNR shift, if compared to AV-HuBERT results. During training different masking strategies were used for data augmentation. With a probability of 50\,\%, we mask a randomly chosen region of each frame with an area of up to 30\,\% of the image and a ratio of 0.3 to 3. Additionally we mask a random sequences of 0 to 10 consecutive frames for each second of input length.
An audio augmentation is performed according to ~\cite{MT_SpecAugmentAS_2019}.

For training we select a batch size of 8 for base.en and 4 for small.en due to the larger model size. For pre-training and the first part of main-training, we set a learningrate of 1e-4. As soon as the training starts to converge, we reduce the learningrate by a factor of 1.58 after each full epoch until the model shows no improvement. For finetuning, including Whisper models, we start after the first part of the main-training with a learning rate of 1e-5 and apply the same learning rate decay strategy. \\
Our approach requires only a single A100 40GB GPU for training, which takes 2.5/2.9 days (wo/w Whisper finetuning) for base.en and 4.6/4.8 days for small.en. AV-HuBERT, on the other hand, requires 32/64 V100 GPUs and takes 2.0/2.4 days for base/large just for pre-training~\cite{AVSR_avhubert_2022}. 

\begin{table*}[ht]
\footnotesize
\centering
\scalebox{0.94}{
\begin{tabular}{rcccccc|ccccc|ccccc|c}
\toprule
\multicolumn{1}{c}{} & \multicolumn{1}{c}{} & \multicolumn{5}{c}{\textbf{Babble SNR [dB]}} & \multicolumn{5}{c}{\textbf{Music+Natural SNR [dB]}} & \multicolumn{5}{c}{\textbf{Sidesp. SNR [dB]}} \\
\cmidrule(rl){3-7} \cmidrule(rl){8-12} \cmidrule(rl){13-17}
 & \textbf{Models} & {-10} & {-5} & {0} & {5} & {10} & {-10} & {-5} & {0} & {5} & {10} & {-10} & {-5} & {0} & {5} & {10} & clean \\
\midrule

(1)  & \begin{tabular}{@{}c@{}}AV-HuBERT base-30h \\ (103\,M)\end{tabular} & \textbf{73.4} & \textbf{45.4} & \textbf{19.8} & 9.5 & 6.1 & 33.2 & 17.8 & 9.8 & 6.7 & 5.4 & 49.4 & 30.2 & 18.6 & 11.2 & 7.5 & 4.3 \\
\\[-0.9em]
(2) & \begin{tabular}{@{}c@{}}Whisper base.en \\ (74\,M)\end{tabular} & 148.6 & 123.9 & 49.0 & 14.6 & 4.7 & 93.0 & 43.6 & 15.2 & 6.1 & 3.7 & 132.4 & 115.6 & 107.0 & 26.7 & 6.5 & 2.5 \\
\\[-0.9em]
(3) & \begin{tabular}{@{}c@{}}Ours + Whisper-base.en \\ (\underline{13\,M} + 74\,M)\end{tabular} & 97.0 & 74.5 & 28.6 & 10.1 & 4.5 & 40.1 & 18.6 & 7.9 & 4.5 & \textbf{3.2} & 55.3 & 56.5 & 32.9 & 14.9 & 5.5 & 2.7 \\
\\[-0.9em]
(4) & \begin{tabular}{@{}c@{}}Ours + Whisper-base.en \\ Finetune Whisper \\ (\underline{13\,M} + \underline{74\,M})\end{tabular} & 82.3 & 53.2 & 22.2 & \textbf{7.0} & \textbf{4.3} & \textbf{28.0} & \textbf{14.2} & \textbf{7.2} & \textbf{4.3} & 3.4 & \textbf{35.4} & \textbf{25.1} & \textbf{17.1} & \textbf{9.2} & \textbf{5.3} & 2.8 \\
\\[-0.9em]
\hline
\\[-0.9em]
(5) & \begin{tabular}{@{}c@{}}AV-HuBERT large-30h \\ (325\,M)\end{tabular} & \textbf{61.4} & \textbf{32.8} & \textbf{12.8} & 6.5 & 4.5 & \textbf{24.9} & 12.3 & 6.5 & 4.6 & 3.9 & 41.6 & 24.9 & 14.4 & \textbf{8.9} & 5.2 & 3.4 \\
\\[-0.9em]
(6) & \begin{tabular}{@{}c@{}}Whisper small.en \\ (244\,M)\end{tabular} & 131.6 & 102.4 & 31.9 & 7.6 & 3.4 & 58.2 & 25.4 & 9.4 & 3.9 & 2.8 & 123.8 & 108.8 & 60.1 & 12.4 & \textbf{3.8} & 2.4 \\
\\[-0.9em]
(7) & \begin{tabular}{@{}c@{}}Ours + Whisper-small.en \\ (\underline{13\,M} + 244\,M)\end{tabular} & 112.0 & 72.7 & 22.9 & 6.7 & 2.9 & 34.3 & 15.3 & 6.8 & 3.5 & \textbf{2.6} & 69.8 & 63.1 & 33.5 & \textbf{8.9} & 3.9 & 2.1 \\
\\[-0.9em]
(8) & \begin{tabular}{@{}c@{}}Ours + Whisper-small.en \\ Finetune Whisper \\ (\underline{13\,M} + \underline{244\,M})\end{tabular} & 100.9 & 55.6 & 16.3 & \textbf{5.2} & \textbf{2.8} & 26.6 & \textbf{12.1} & \textbf{4.9} & \textbf{3.0} & \textbf{2.6} & \textbf{29.9} & \textbf{20.0} & \textbf{13.8} & 10.3 & 6.5 & 2.3 \\

\bottomrule
\end{tabular}
}
\caption{WER [\%] of Whisper (audio-only), AV-HuBERT (audio-video) and our approach in combination with Whisper (audio-video) on the LRS3 dataset. Two groups according to model sizes. Values in brackets denote the number of model parameter while the underlining identifies the trainable parameters. Bold values mark best result for each group, noise category and noise level.}
\label{tab:res}
\end{table*}

\section{Results}
\label{sec:results}

\tablename~\ref{tab:res} shows the results for different combinations of our fusion module, the Whisper ASR and AV-HuBERT models. For comparability, we divide the models in 2 groups, according to the number of parameters. We basically compare the fusion module and Whisper base.en to AV-HuBERT base as well as our module and Whisper small.en to AV-HuBERT large. The number of parameters for each variant is noted in brackets, below each model name in Table \ref{tab:res}.  
Our approach, combining Whisper and the fusion module, contains 16\,\% less parameter for base.en and 21\,\% less parameter for small.en than the corresponding AV-HuBERT models. For all variants, we select the same input data from the official testset of LRS3-Ted dataset. We analyse the results over different SNR values (-10, -5, 0, 5, 10)\,dB and for the noise categories babble, music, natural and sidespeaker. Due to the similarity of the noise types and the results, we combine the noise categories music and natural into one result group music+natural. For this purpose, the results for both categories are computed separately. For analysis we calculate the average values. In addition to the AVSR models, we analyse the accuracy of the Whisper baseline models for the same noisy audio inputs. We compare and evaluate all models using the Word Error Rate (WER) metric.\\
\\[-0.7em]
\textbf{Whisper baseline models} \\
Although Whisper provides high robustness to ambient noise compared to other ASR models, an exponential increase in WER is observed with increasing ambient noise (decreasing SNR values) in rows 2 and 6 of \tablename~\ref{tab:res}. As expected, the weakness towards spoken noises like babble or sidespeaker is significantly higher compared to other noise categories. This is explainable, since the model is no longer capable to determine which speech signal to focus on, especially for SNR values lower than 0\,dB. For low ambient noise, at SNR values greater or equal to +10\,dB and clean inputs Whisper performs better than AV-HuBERT (rows 1 and 5). This is related to the training with enormous amounts of mostly clean speech data. \\
\\[-0.7em]
\textbf{Fusion module with frozen Whisper model} \\
At first stage we train our fusion module adaptive to the Whisper models base.en and small.en (ours + Whisper base.en / small.en). 
The ASR models remain frozen, so the training is limited to optimize the fusion module with 13\,M parameters with the aim to enhance the ASR inputs and therefore improve the ASR outputs.
The results in \tablename~\ref{tab:res}, rows 3 and 7, show a significant improvement compared to the audio only Whisper baseline models across all noise categories and allmost all SNR values. We achieve an average WER reduction of about 30\,\% for base.en and about 20\,\% for small.en. The decrease in improvement for the larger Whisper model (small.en) could be the result of the lower parameter ratio. While Whisper base.en contains 74\,M, small.en contains 244\,M parameter, which increases the complexity of information processing and hypothesis for small.en. We apply the same fusion module size with 13\,M parameter to both ASR models, which leads to a decreased parameter ratio for the combination with small.en. This may also affect the increased WER values at babble noise for SNR -5\,dB and -10\,dB and sidespeaker noise for SNR 0\,dB to -10\,dB for small.en compared to base.en.
Compared to AV-HuBERT (rows 1 and 5), we achieve an improvement in the recognition rate across all noise categories with low ambient noise, high SNR value. This is also influenced by the very high recognition rate of Whisper for clean audio signals, which is further improved by the fusion module and extended to the SNR range between +10\,dB and +5\,dB. At low SNR values, AV-HuBERT shows significantly higher recognition rates compared to our proposed apprach without finetuning. This is particularly true for both speech noise categories babble and sidespeaker.  \\
\\[-0.7em]
\textbf{Fusion module with finetuned Whisper model} \\
Finally we unfreeze both Whisper models and finetune the combinations with our fusion module. The results in Table \ref{tab:res}, rows 4 and 8, show a further improvement compared to the variants with frozen Whisper models in rows 3 and 7 and the Whisper baseline model in rows 2 und 6 across almost all noise types and SNR values, except small.en for sidespeaker noise with SNR values greater or equal to +5\,dB. With respect to the Whisper baseline models, this is most evident for the categories sidespeaker and music+natural, where we achieve a reduction in word error rate of up to 75\,\% after finetuning.

In comparison to the results of AV-HuBERT in rows 1 and 5, our approach achieves an improvement in the categories music+natural and sidespeaker across almost all SNR values, except the combination with small.en in row 8 for the category music+natural at SNR -10\,dB and the category sidespeaker at SNR +5\,dB and +10\,dB.
On average across both model sizes and all SNR values, our approach improves the results for category sidespeaker by 11.2\,\% and music+natural by 22.4\,\%.

In contrast, our approach suffers from clear weaknesses in the babble category, where we achieve an improvement for SNR +5\,dB and +10\,dB over both model sizes, but perform worse than AV-HuBERT from SNR 0 and below. For this noise category our results show an increase in WER of 8.8\,\% on average across all model sizes and SNR values compared to AV-HuBERT. In particular, the combination with small.en, in row 8, shows partly almost 70\,\% higher WER values. Different factors can be responsible for this behavior. First, our approach that combines the fusion module with Whisper consists of 16\,\% less trainable parameter for base.en and 21\,\% less for small.en than the corresponding AV-HuBERT models. Additionally, the focus during training lies on a broad range of equally distributed SNR values in the range [-20\,dB, +50\,dB]. It may be possible to improve the results for low SNR values by shifting the focus towards these values during training. In general, babble noise is problematic for all model variants. An explanation might be that babble noise contains a very high variability so that the number of training examples, also compared to the other noise categories, is too small to achieve a better generalisation.
\\[1.0em]

\section{Conclusion}
\label{sec:conclusion}

We presented a strategy to improve the results of ASR models under noisy environments by using an adaptive trainable AV fusion module. Without ASR finetuning the fusion module achieves an improved performance compared to the baseline ASR models, but does not reach the performance of AVSR models, like state of the art approach AV-HuBERT which is trained from scratch. 
By finetuning the overall model, we outperform the results of AV-HuBERT for the noise categories sidespeaker by 11.2\,\% and music+natural by 22.4\,\% on average for a broad range of SNR values.
For babble noise, our fusion module achieves better results than AV-HuBERT at low noise levels, but cannot keep up with AV-HuBERT at higher noise levels, which lead to a deterioration of 8.8\,\% on average. Overall, this leads to an average improvement of 8.3\,\% across all noise categories and model sizes.
In sum, our approach can keep up with the performance of current state of the art approaches to AVSR, like AV-HuBERT, despite 21\,\% lower number of parameter compared to AV-HuBERT large and the use of significantly lower computational resources. 

Yet, our approach offers some starting points for further performance improvement. A comparison of the combination with different ASR model sizes has shown that the WER improvement decreases with an increasing ASR model size if the fusion module size is kept constant. This could be related to the resulting change in parameter ratio between fusion module and ASR model. 
Increasing the number of fusion module parameter in combination with larger ASR models could lead to a better adaptability to the increased processing and hypothesis complexity.
Furthermore, our experiments have shown that for the noise categories sidespeaker, music and natural, our approach achieves significantly better results than AV-HuBERT, while for babble noise in combination with low SNR values AV-HuBERT is superior. A change of the training sample distribution and SNR value distribution could be beneficial, to provide a higher focus on babble noise and lower SNR values. That may lead to an improvement for this category and SNR values with acceptable minor degredations for other noise categories and noise levels.

\newpage
\bibliographystyle{IEEEbib}
\bibliography{strings,refs}

\end{document}